\documentclass[twocolumn,showpacs,preprintnumbers,amsmath,amssymb,prb]{revtex4}

\usepackage{graphicx}
\usepackage{dcolumn}
\usepackage{bm}

\begin{document}
\draft
\title{Raman characteristic peaks induced by the topological defects of Carbon
Nanotube Intramolecular Junction}

\author{Gang Wu}
\altaffiliation[Present address: ]{Department of Physics, National
University of Singapore, Singapore 117542}
\email{wugaxp@gmail.com}

\author{Jinming Dong}
\email[Corresponding author. Email address: ]{jdong@nju.edu.cn}

\affiliation{National Laboratory of Solid State Microstructures and
Department of Physics, Nanjing University, Nanjing 210093, P. R.
China}

\date{\today}

\begin{abstract}

The vibrational modes of some single wall carbon nanotube (SWNT)
intramolecular junctions (IMJs) have been calculated using the
newest Brenner reactive empirical bond order (REBO) potential, based
upon which their nonresonant Raman spectra have been further
calculated using the empirical bond polarizability model. It is
found that the Raman peaks induced by pentagon defects lie out of
the $G$-band of the SWNTs, so the high-frequency part of the Raman
spectra of the SWNT IMJs can be used to determine experimentally
their detailed geometrical structures. Also, the intensity of the
Raman spectra has a close relation with the number of pentagon
defects in the SWNT IMJs. Following the Descartes-Euler Polyhedral
Formula (DEPF), the number of heptagon defects in the SWNT IMJs can
also be determined. The first-principle calculations are also
performed, verifying the results obtained by the REBO potential. The
$G$ band width of the SWNT IMJ can reflect the length of its
transition region between the pentagon and heptagon rings.

\end{abstract}

\pacs {63.22.+m, 78.30.Na, 61.46.Fg}

\maketitle

\section{Introduction}

Single-wall carbon nanotube (SWNT) is a kind of nanoscale molecule
obtained by wrapping a graphene sheet into a seamless cylinder.
Because of its remarkable electronic properties \cite{r1}, SWNT is
expected to play an important role in the nanoscale electronics,
e.g., the field-effect transistor (FET) based logic devices
\cite{r2, r3}, and in applications of nanophotonics \cite{r4}. One
of the most important structures within these devices is the SWNT
intramolecular junctions (IMJs), which are formed by introducing the
pentagon-heptagon rings in them. Different chiral SWNTs can be
connected through the IMJs and form the so-called metallic-metallic
(M-M), metallic-semiconductor (M-S), or semiconductor-semiconductor
(S-S) junctions. It is well-known that the electronic and optical
properties of the SWNT IMJs have a close relationship with their
geometrical or topological characteristics. A lot of theoretical
studies on the electronic \cite{r5, r6, r7, r8, r9, r10, r11, r12,
r13} and optical \cite{r14} properties of the SWNT IMJs have been
carried out to get indirectly the detailed information on their
topological structures. At the same time, many experiments have been
performed to study the transport properties of the SWNT IMJs
\cite{r15, r16, r17, r18}. Recently, the direct Scanning Tunneling
Microscope (STM) imaging and the measurements of the density of
states (DOS) have provided the best direct evidences of the SWNT
IMJs \cite{r19, r20}. On the other hand, some efforts have been made
to investigate the vibrational properties of the SWNT IMJs, e.g.,
the measurement of the radial breathing mode (RBM) by combining
inelastic electron tunneling spectroscopy and STM method \cite{r21},
and the Raman spectra by the confocal Raman spectral imaging
\cite{r22}. Although the vibrational property obviously is very
important for determination of the SWNT IMJs' structures, yet no
theoretical study on their Raman spectra has been reported to date.

Beyond the traditional force-constant method \cite{r23}, the
state-of-the-art first-principles methods have already been used
to calculate more accurately the phonon dispersion curves
\cite{r24, r25, r26, r27, r28}, offering a unique technique to
investigate the vibrational properties of the nanotubes or
nanoropes. But for the very low symmetrical systems containing
many atoms, such as SWNT IMJs, the first-principles method demands
too intensive calculations. So, in this paper, we use the highly
accurate empirical potential to calculate the vibrational
properties of some SWNT IMJs. Here, the empirical potential is
chosen to be the second-generation reactive empirical bond order
(REBO) potential \cite{r29}, which is the newest version of the
Tersoff-Brenner type potential, combining advantages of the two
sets of parameters in the earlier version \cite{r30}. The first
set of parameters (Brenner I) underestimates the isotropic elastic
constants, while the second set of parameters (Brenner II)
overestimates the interatomic distance. The new REBO potential
contains improved analytic functions and an extended database,
offering a significantly better description of bond energies, bond
lengths, and force constants for the hydrocarbon molecules, as
well as elastic properties, interstitial defect energies, and
surface energies for the diamond \cite{r29}. As a result, it
should have a satisfying accuracy for describing the vibrational
properties of the SWNT IMJs.

After having obtained the vibrational modes of the SWNT IMJs, the
nonresonant Raman intensity is further calculated by the empirical
bond polarizability model \cite{r31, r28}. And for their Raman
spectra, we will pay more attention to those characteristic Raman
peaks caused by pentagon or heptagon defects because they are most
important for investigating the local structures of the SWNT IMJs.

In the following section, the REBO potential is first introduced and
its ability to describe the vibrational properties of SWNTs is
discussed. Then, the vibrational modes of some SWNT IMJs are
calculated using the REBO potential. It is found that the
characteristic Raman peaks caused by the pentagon defects lie out of
the highest Raman peak of the SWNTs, which so can be regarded as an
indicator of the pentagon defects. In the later part of the paper,
we will show that the heights of these peaks have some relationships
with the number of the pentagon defects. On the contrary, the
heptagon defects can only cause weak localized states below the $G$
band of the SWNTs, which can be hardly observed in experiments.
Furthermore, the junction length is shown to be able to influence
the $G$ band width of SWNT IMJ. In other words, the longer length of
the transition region between the pentagon and heptagon rings means
the wider distribution of the $G$ band. The first-principles
calculations are carried out on the graphene sheet with Stone-Wales
defect on it in order to further verify our results quantitatively.
Finally, this paper ends with some concluding remarks.

\section{Numerical methods }

\begin{table}[htbp]
\caption{\label{table1}The Raman frequencies of (10, 10), (17, 0)
and (5, 5) SWNTs. The unit is cm$^{-1}$. Three different methods,
i.e., the REBO potential, tight-binding method and LDA method, are
used to obtain the dynamic matrix. The TB results are taken from
Ref. \onlinecite{r32}. The Raman mode around 900 cm$^{-1}$ is
$E_{2g}$ in armchair tube and $E_{1g}$ in zigzag tube, so in this
table, it is labeled as $E_{2(1)g}$.}
\begin{ruledtabular}
\begin{tabular}{cccccccccc}
{}& \multicolumn{3}{c}{(10, 10)} & \multicolumn{3}{c}{(17, 0)} &
\multicolumn{3}{c}{(5, 5)}  \\
\cline{2-4} \cline{5-7} \cline{8-10}
 &
REBO& TB& LDA& REBO& TB& LDA& REBO& TB&
LDA \\
\hline $E_{2g}$ & 19& 16& 28& 28& 17& 23& 58& &
98 \\
\hline $E_{1g}$ & 94& 91& 102& 87& 93& 110& 212& &
202 \\
\hline $A_{1g}$ & 151& 157& 172& 162& 160& 176& 306& &
355 \\
\hline $E_{2g}$ & 336& 334& 375& 353& 351& 376& 664& &
710 \\
\hline $E_{2(1)g}$ & 930& 942& 879& 951& 953& 874& 873& &
837 \\
\hline $E_{1g}$ & 1650& 1673& 1584& 1689& 1685& 1601& 1806& &
1569 \\
\hline $A_{1g}$ & 1662& 1679& 1588& 1697& 1672& 1600& 1727& &
1589 \\
\hline $E_{2g}$ & 1691& 1700& 1616& 1626& 1658& 1568& 1668& & 1583
\end{tabular}
\end{ruledtabular}
\label{tab1}
\end{table}

Firstly, the phonon dispersion relationship and vibrational DOSs of
armchair (10, 10) and zigzag (17, 0) tubes are calculated to
demonstrate the accuracy of the REBO potential. Obtained results are
shown in Figure 1. Correspondingly, the frequencies of the Raman
active modes are presented in Table I. For comparison, the Raman
frequencies calculated by both the tight-binding (TB) and the
first-principles methods are also given. Here, the TB results are
taken from Ref. \onlinecite{r32}.

\begin{figure}
\includegraphics[width=3.24in,height=2.53in]{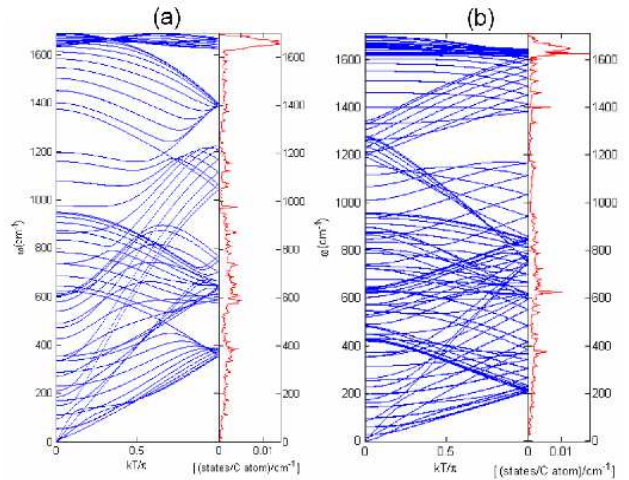}
\caption{\label{fig1} (color online) The phonon dispersion
relationship and vibrational density of states of armchair (10, 10)
and zigzag (17, 0) SWNTs, calculated by REBO potential. Left part is
for (10, 10) tube, and right one is for (17, 0) tube.}
\end{figure}

The first-principles calculations are made as follows. A supercell geometry was adopted so that the SWNTs are
aligned in a hexagonal array with nearby SWNT center distance of 25 {\AA}, which is found to be larger enough to
prevent the tube-tube interactions. One unit cell of SWNT is sufficient to calculate accurately the phonon modes
at the $\Gamma $ point. The $k$ points sampling in the reciprocal space is a uniform grid along the nanotube
axis with the maximum spacing between $k$ points being 0.03 {\AA}$^{ - 1}$ and the Gaussian smearing width is
0.03 eV. A plane-wave cutoff of 400 eV is used to obtain reliable results. After structure relaxation on both of
the lattice constant along the tube axis and the atomic positions, the optimal structure is obtained when the
residual forces acting on all the atoms were less than 0.01 eV/{\AA}. Our \textit{ab initio} calculations were
performed using highly accurate projected augmented wave (PAW) method \cite{r33}, implemented in the Vienna
\textit{ab initio} simulation package (VASP) \cite{r34}, which is based on the density-functional theory in the
local-density approximation (LDA).

From Fig. 1 and Table I, one can conclude that the REBO potential can give
reasonable lower Raman frequencies. As for the higher Raman modes, the REBO
potential underestimates the slope of dispersion relationship (or the group
velocity) and overestimates the frequencies. But its results agree well with
those obtained by the TB method even quantitatively. In a word, the REBO
method can be used to calculate the Raman modes of the SWNTs, especially for
the tubes with larger radii.

Next, we will introduce the construction method of the SWNT IMJs used in
this work.

The surface of SWNT and SWNT IMJs obeys the Descartes-Euler Polyhedral
Formula (DEPF)\textbf{,} i.e., for the open-ending SWNT or SWNT IMJ, the
following equation should be obeyed,

\begin{equation}
\label{eq1} V - E + F = 2(1 - G).
\end{equation}

Here, $V$, $E$, and $F$ are the number of vertices, edges and faces, respectively, and $\chi = 2\left( {1 - G}
\right)$ is the Euler characteristic. $G$ is the genus, which is always equal to $1$ for SWNT and SWNT IMJ with
only one hole. For perfect SWNTs, $E = \frac{3}{2}V$, $F = \frac{V}{2}$, $G = 1$, and Eq. 1 is valid. If we
further let $N_5 $ and $N_7 $ denote the number of pentagon and heptagon rings, respectively, then Eq. 1 means
that $N_5 - N_7 = 0$ for any SWNT IMJ. In other words, the number of pentagon rings should equal to the number
of heptagon rings.

In Ref. \cite{r35}, a universal algorithm is carried out to connect
arbitrary two SWNTs with one pair of pentagon-heptagon rings.
Obviously, the structures generated by this method have the minimum
defect energy. So, in this work, we also use this method to generate
the SWNT IMJ structures.

\section{Numerical results and discussions}

Now, we firstly calculate the vibrational spectrum of a SWNT IMJ
constructed by the (10, 10) and (17, 0) tubes with almost the same
radii, whose structures are shown in Fig. 2. In order to eliminate
the effect of dangling bonds, we further connect two same (10,
10)-(17, 0) SWNT IMJs to build a periodic structure shown in the
right panel of Fig. 2, which is studied in this work. Then a
structural optimization is performed to prevent the appearance of
imaginary frequency, i.e., the soft mode.

\begin{figure}
\includegraphics[width=3.24in,height=1.33in]{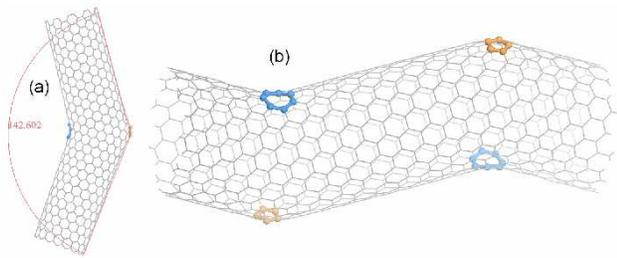}
\caption{\label{fig2} (color online) (10, 10)-(17, 0) SWNT IMJ. The
optimized structure is shown in the left panel and the periodic
structure is shown in right panel. Only one pair of
pentagon-heptagon rings is used to connect the different tubes.}
\end{figure}

In Fig. 3, the nonresonant Raman spectra of some structures are presented. Because the low frequency $R$ band
and the high frequency $G$ band are mostly interested in experiment, only those two frequency ranges, i.e.,
0-400 cm$^{ - 1}$ and 1400-2100 cm$^{ - 1}$, are shown in Fig. 3.

\begin{figure}
\includegraphics[width=3.24in,height=1.33in]{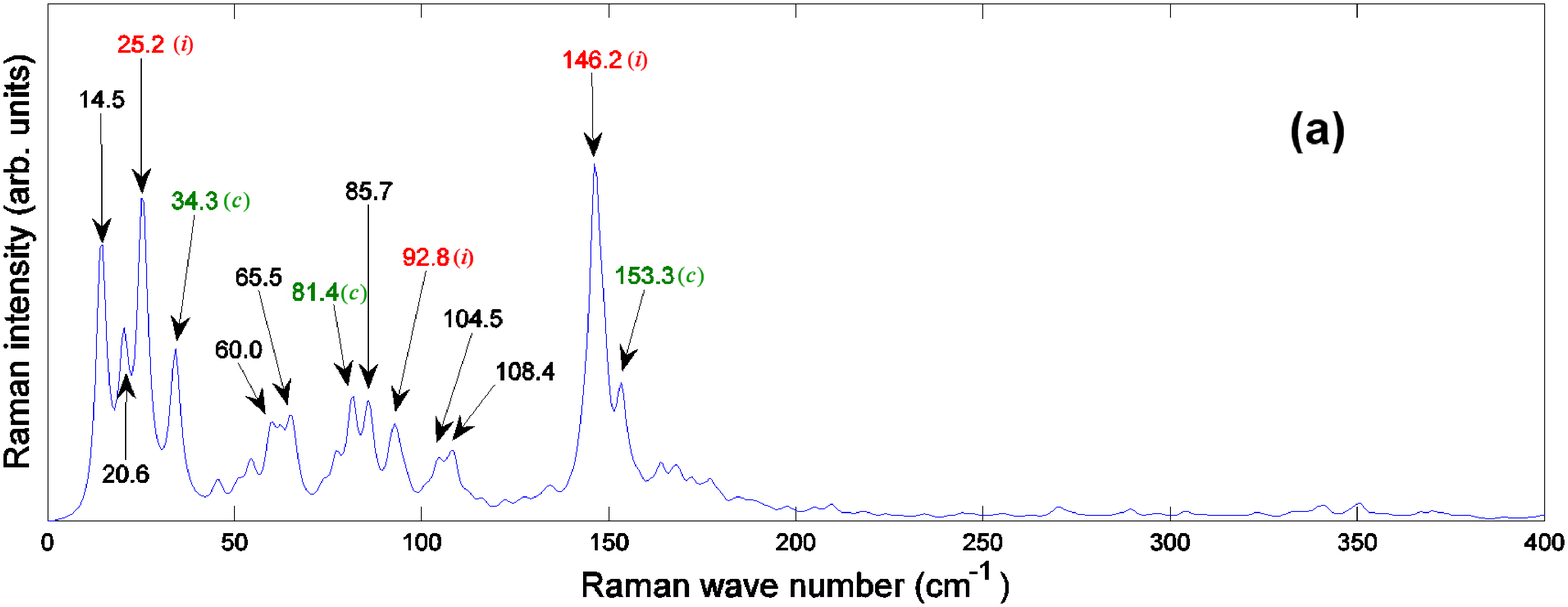}
\includegraphics[width=3.24in,height=1.33in]{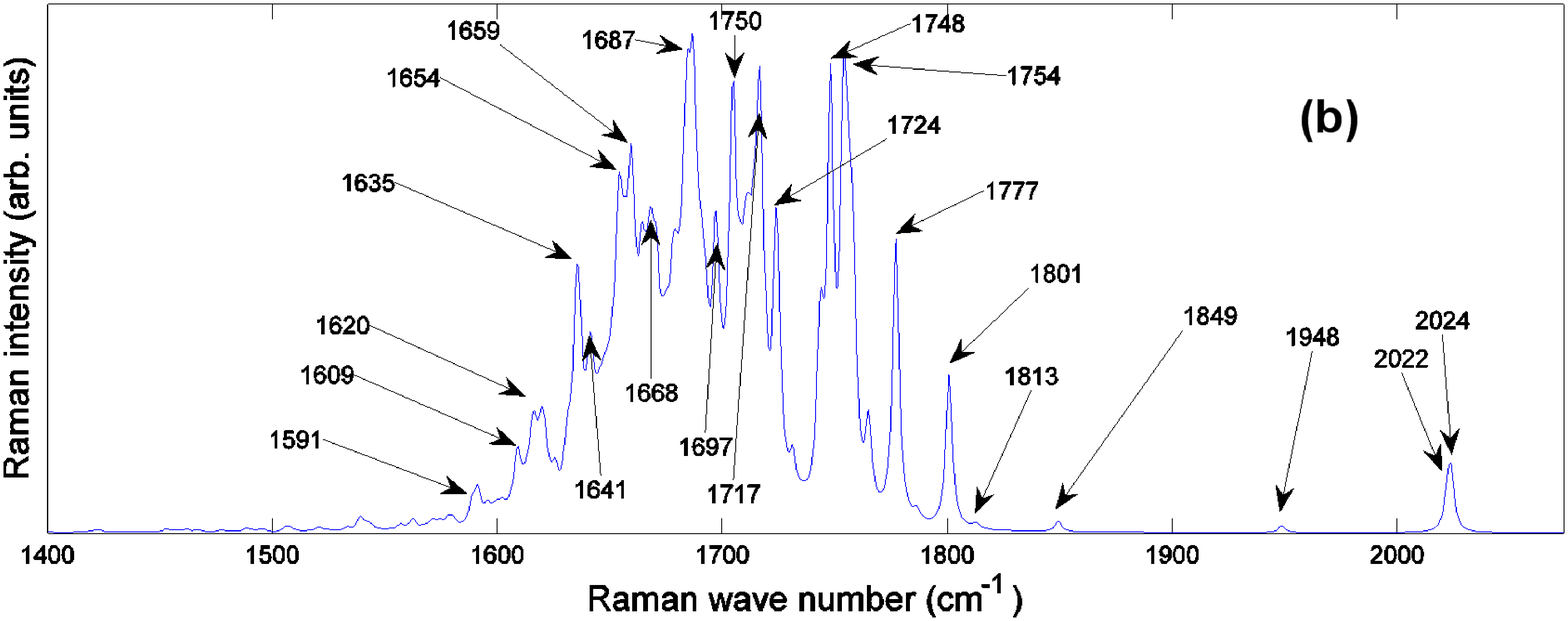}
\includegraphics[width=3.24in,height=1.33in]{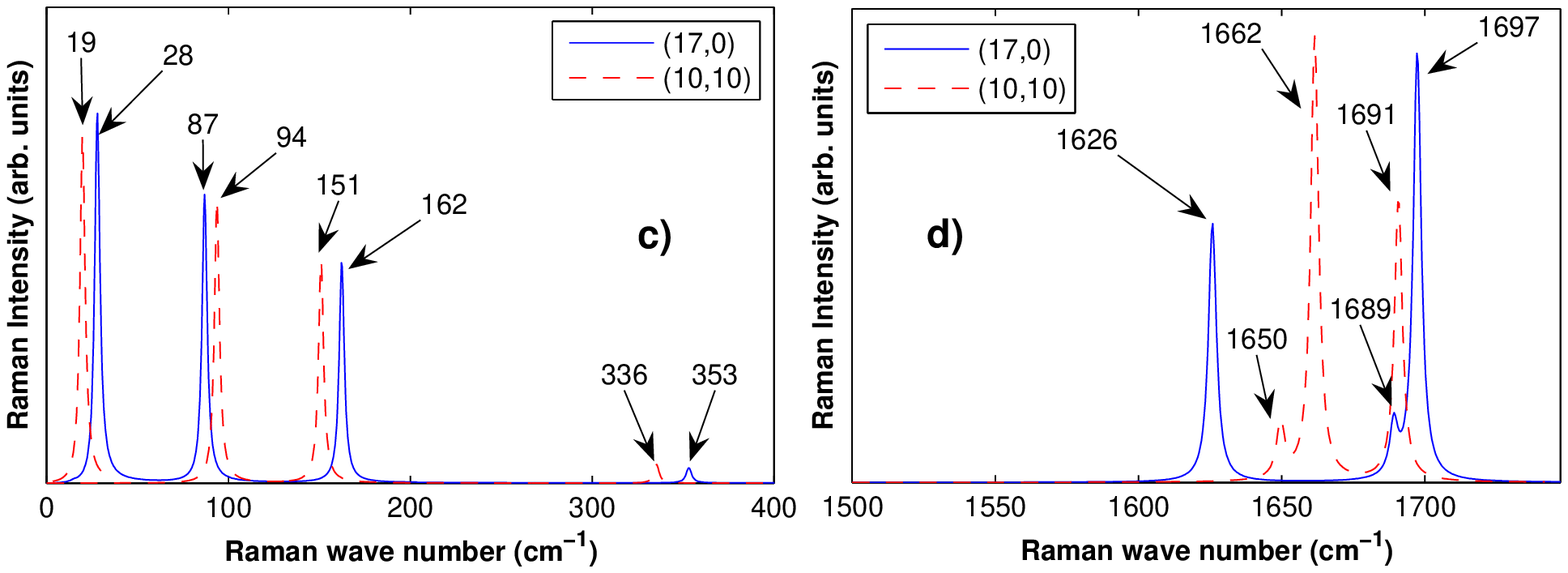}
\caption{\label{fig3} (color online) The nonresonant Raman
spectrum of the (10, 10)-(17, 0) SWNT IMJ. (a) Low frequency part.
(b) High frequency part. (c) and (d) show the low and high
frequency parts of the Raman spectra of (10, 10) and (17, 0)
tubes, respectively. In Fig. 3(a), the letters of (i) and (c)
indicate that the modes come from the in-phase and counter-phase
combinations of the original Raman modes in the perfect SWNTs,
respectively.}
\end{figure}

Comparing Fig. 3a and 3c, it is clearly seen that many small peaks appear in the low frequency part. Some of
them come from the original SWNT components, and their frequency numbers are affixed by letters of ($i)$ or
($c)$, indicating, respectively, the corresponding vibration mode comes from the in-phase or counter-phase
combination of the Raman modes in two SWNTs. For example, the peak at 25.2 cm$^{ - 1}$ roots in the in-phase
combination of the $E_{2g}$ modes of (10, 10) and (17, 0) tubes, but the peak at 34.3 cm$^{ - 1}$ comes from
their counter-phase combination. Their atomic vibrations have been shown in Fig. 4, from which it is clearly
seen that the atomic motions on the (10, 10) and (17, 0) tubes are in the same direction for 25.2 cm$^{ - 1}$
mode, whereas they are in the reverse directions for 34.3 cm$^{ - 1}$ mode.

\begin{figure}
\includegraphics[width=3.24in,height=1.47in]{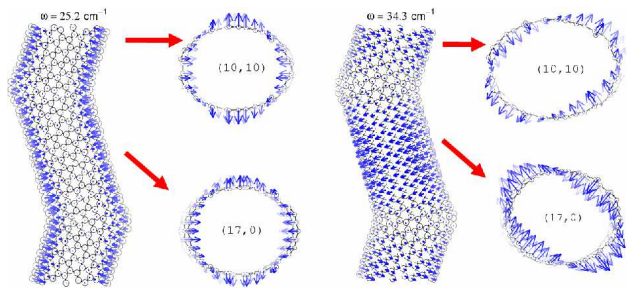}
\caption{\label{fig4} (Color online) The atomic vibrations of the
Raman active modes at frequencies of 25.2 and 34.3 cm$^{- 1})$,
which come from the combinations of the basic Raman active modes
of the (10, 10) and (17, 0) SWNTs.}
\end{figure}

Some other peaks in the low frequency part, e.g., the peaks at 14.5, 20.6, 60.0, 65.5, 85.7 and 108.4 cm$^{ -
1}$, can not find their counterparts in the Raman spectra of the original SWNTs. Some of them come from the
symmetry breaking of the perfect nanotube, and some others come from the vibration of the connecting part of the
IMJ. For example, the atomic vibrations of 65.5, 85.7 and 108.4 cm$^{ - 1}$ modes are shown in Fig. 5.
Obviously, the atomic motions near the defect parts are larger than those on the tube. But, most probably, it is
practically difficult to observe these peaks because the most intensive low frequency peak found in experiments
is the $R$ band caused by RBM mode.

\begin{figure}
\includegraphics[width=3.24in,height=5.84in]{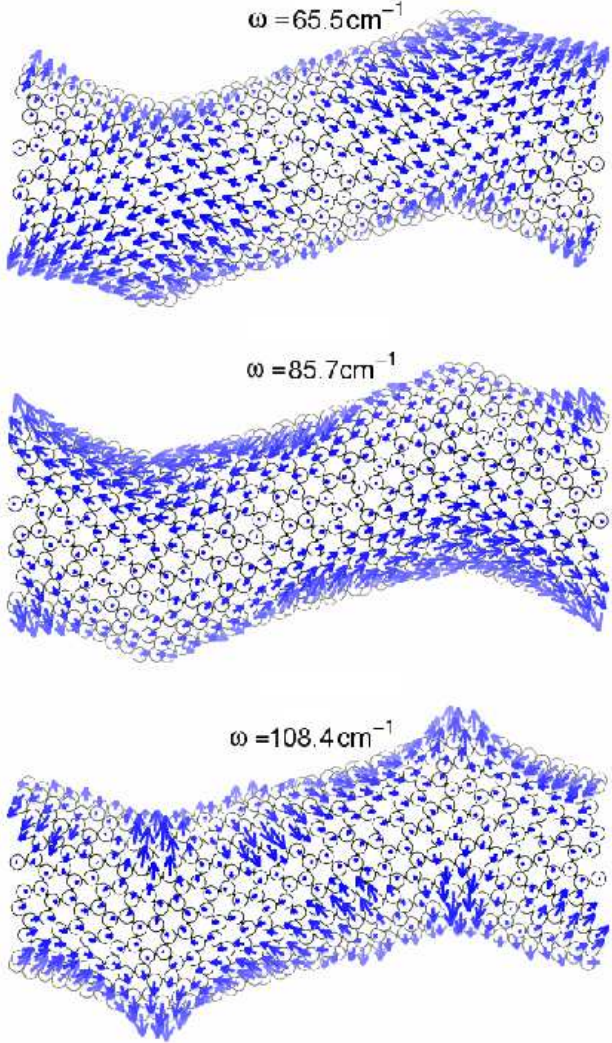}
\caption{\label{fig5} (Color online) The atomic vibrations of the
Raman active modes at frequencies of 65.5, 85.7 and 108.4
cm$^{-1})$, which come from the atomic vibrations of the junction
part.}
\end{figure}

As for the high frequency part, it can be found from Fig. 3b and 3d that many peaks appear in the $G$ band
(1600-1800 cm$^{ - 1})$. A superposition of these peaks forms a widely distributed $G$ band, which could not be
regarded as a simple superposition of those $G$ bands from the original SWNTs. This phenomenon has been observed
in the experimental result of Ref. \onlinecite{r21}, making also the detailed analysis on the $G$ band of SWNT
IMJ very complicated and hardly to be performed.

On the other hand, in the range out of $G$ band, i.e., in a range of about
1800-2050 cm$^{ - 1}$, some other Raman peaks can be found, whose modes are
much simpler than the $G$ band, and so may offer some useful information about
the topological structures of SWNT IMJs.

\begin{figure}
\includegraphics[width=3.24in,height=5.4in]{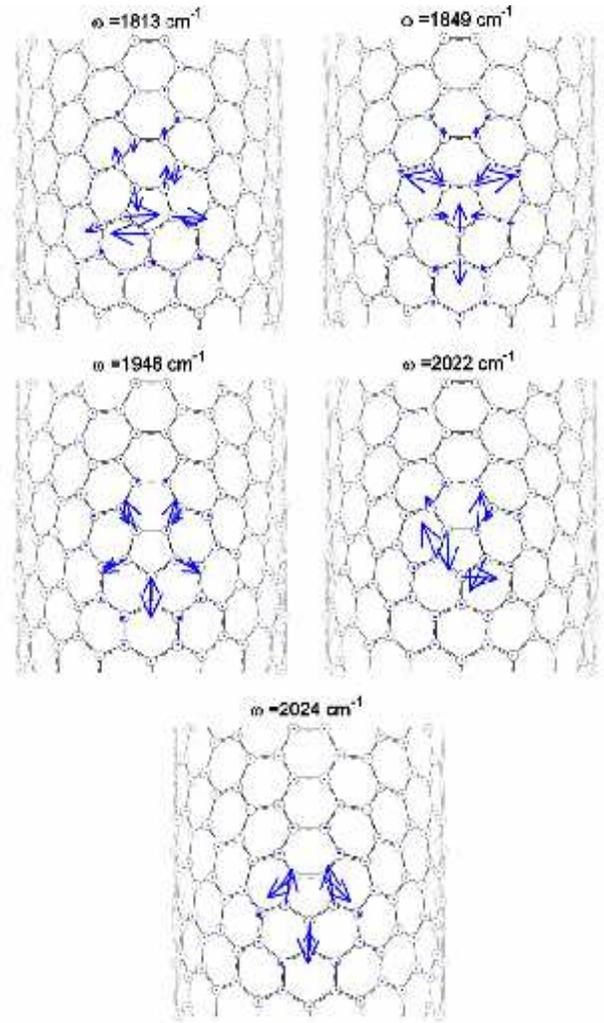}
\caption{\label{fig6} (Color online) The atomic vibrations of the Raman active modes beyond 1800 cm$^{ - 1}$ for
the (10, 10)-(17, 0) SWNT IMJ.}
\end{figure}

In order to understand the origin of these modes, their atomic
vibrations are presented in Fig. 6. Obviously, these modes have a
common character that their vibrational amplitudes are the largest
on the pentagon ring, and become smaller and smaller on the other
atoms of the tube surface when the atoms lie from the pentagon ring
farther and farther. In other words, these modes are defect states
caused by the pentagon ring. So, the pentagon can be regarded as a
``topological impurity'' in the perfect SWNT. Here, it is found that
the bond length on the hexagon rings is about 1.42 {\AA}, while the
mean bond lengths on the pentagon and heptagon rings are about 1.37
and 1.45 {\AA}, respectively. This fact means that the force
constants on the pentagon (heptagon) defects should be larger
(smaller) relative to those on the hexagon rings. Also, the distance
between the pentagon and heptagon defects is quite large so that
their interaction can be neglected. So, we can simply use the one
dimensional (1D) single impurity model in the classical solid state
theory \cite{r36} to discuss qualitatively the vibrational property
of here's (10, 10)-(17, 0) SWNT IMJ, and find that the frequency of
the ``topological impurity'' state could be written as

\begin{equation}
\label{eq2} \omega ^2 = \omega _M^2 \frac{f^2}{f_0 \left( {2f - f_0 } \right)} = \omega _M^2 \frac{\left( {f
\mathord{\left/ {\vphantom {f {f_0 }}} \right. \kern-\nulldelimiterspace} {f_0 }} \right)^2}{2\left( {f
\mathord{\left/ {\vphantom {f {f_0 }}} \right. \kern-\nulldelimiterspace} {f_0 }} \right) - 1}.
\end{equation}

Here, $f$ and $f_0 $ are the force constant of the ``topological impurity'' and the 1D perfect lattice,
respectively, with its maximum vibrational frequency of $\omega _M (\omega _M^2 = {4f_0 } \mathord{\left/
{\vphantom {{4f_0 } M}} \right. \kern-\nulldelimiterspace} M)$. And $M$ is the atom mass. Eq. 2 is valid when $f
> f_0 $, which evidently corresponds to the situation of the pentagon defects. From Eq. 2, one can find that the
defect frequency $\omega $ is higher than the maximum frequency of the perfect lattice$\omega _M $, which is
also well consistent with our numerical result. Furthermore, if we use the technical term in the traditional
mass-impurity model in the classical solid state theory \cite{r36}, one may consider the pentagon defect to be
the ``light-mass impurity''. But the frequency of the localized defect state can not be simply obtained from Eq.
2 because it is difficult to make a direct link between the simple model and the realistic structure.

Since the pentagon defect can cause localized defect state, then how about the heptagon defect? After searching
for the whole Raman spectrum, we find that it is very hard to induce the localized states on the heptagon
defect. The atomic vibrations of its two strongest defect states are shown in Fig. 7, which is found to be much
weaker than those in pentagon defect states. Correspondingly, the Raman intensities of these modes are much
weaker too. This phenomenon can be understood as follows. Similar to the case of pentagon defect, the heptagon
defect can be regarded as the ``heavy impurity'' in the traditional mass-impurity model, which mainly causes the
phase shift of the vibrational modes. And the frequencies of the localized modes caused by the ``heavy
impurity'' are all lower than the maximum vibrational frequency of the perfect lattice, which is also consistent
with our numerical result.

\begin{figure}
\includegraphics[width=3.24in,height=1.80in]{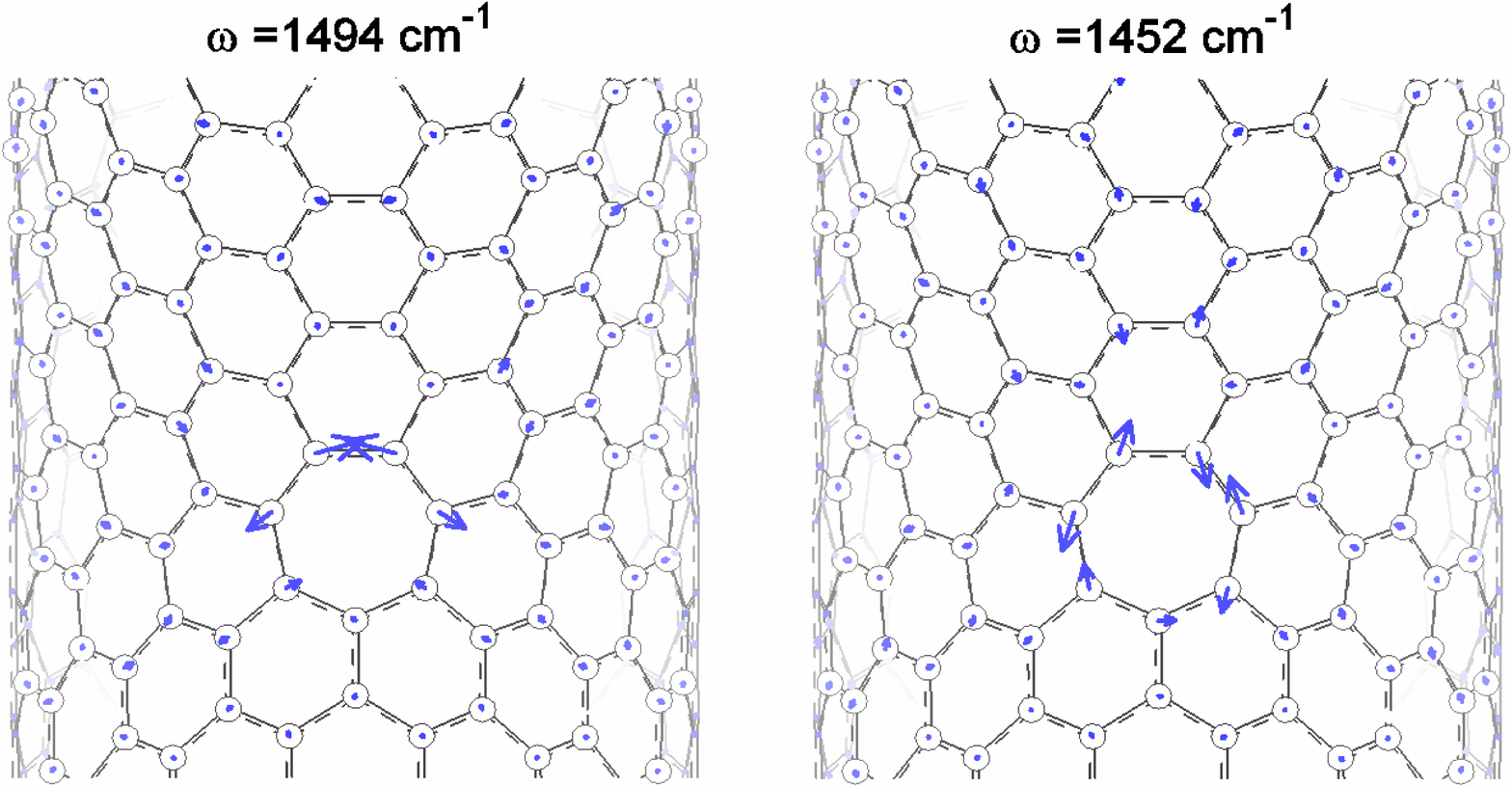}
\caption{\label{fig7} (Color online) The atomic vibrations of the localized states on the heptagon defect of the
(10, 10)-(17, 0) SWNT IMJ. Here only the strongest two states are shown.}
\end{figure}

To further validate our finding about the localized states, the
first-principles calculation is made to study the vibrational properties of
the pentagon-heptagon defects. But as indicated, the first-principles
calculation on the SWNT IMJs is too expensive, and so is made on another
substitute structure, i.e., a graphene sheet with a Stone-Wales (SW) defect
on it. The SW defect is two pairs of pentagon-heptagon defects got together,
and can be generated by rotating one bond in the graphene sheet by 90\r{ }.
Now, the K-points sampling in the reciprocal space is a uniform grid in the
graphene sheet and the maximum spacing between k points is still 0.03
{\AA}$^{ - 1}$.

Firstly, the nonzero vibrational modes in a perfect graphene are
calculated by the first-principles method, which is found to be
1591.38, 1591.14, and 896.24 cm$^{ - 1}$, and similar to the results
in Ref. \onlinecite{r27}. Then, we have performed the calculation on
a periodical (6$\times $6) graphene unit cell with a SW defect in
its center. The distance between nearby graphene sheets is fixed to
be 15 {\AA}, which is large enough to neglect the interaction
between the sheets. Structural optimization is also made to prevent
the soft modes. In such a structure, the distance between the SW
defects is 14.5 {\AA}, which is also large enough to neglect the
interaction between neighbor SW defects.

\begin{figure}
\includegraphics[width=3.24in,height=3.0in]{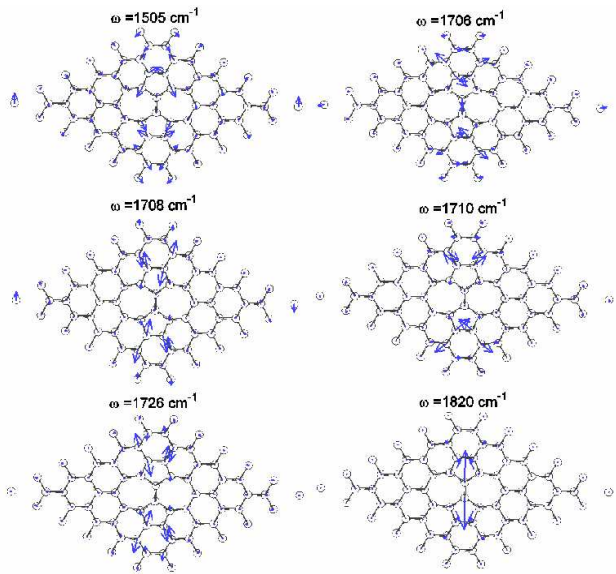}
\caption{\label{fig8} (Color online) Some localized states caused by pentagon defects on a graphene.}
\end{figure}

\begin{figure}
\includegraphics[width=3.24in,height=3.0in]{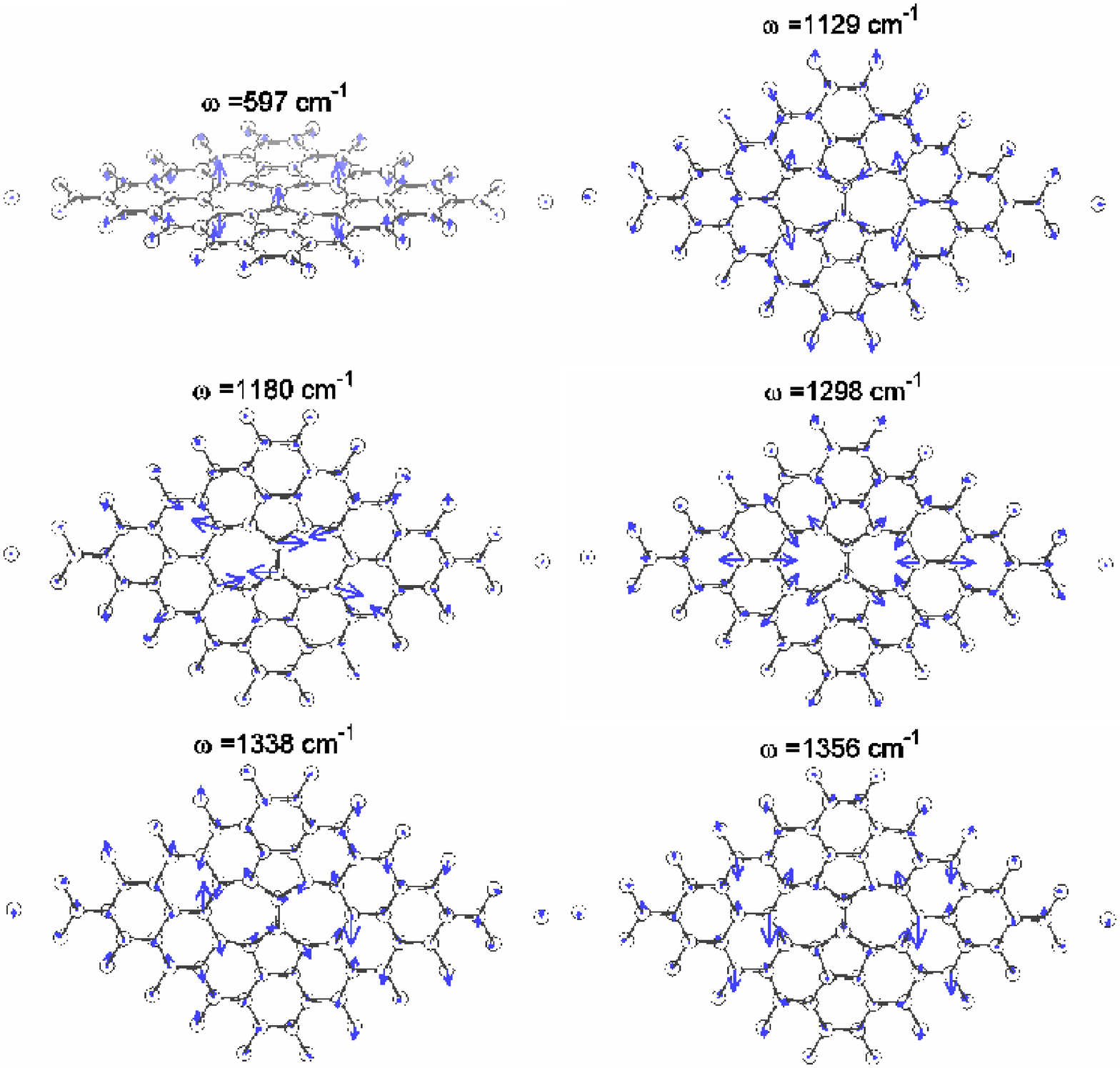}
\caption{\label{fig9} (Color online) Some localized states caused by heptagon defects on a graphene.}
\end{figure}

\begin{figure}
\includegraphics[width=3.24in,height=1.0in]{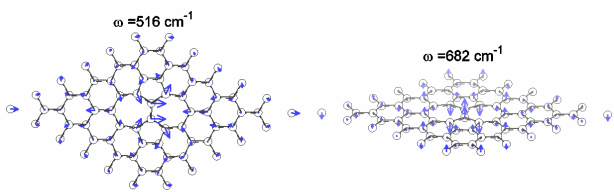}
\caption{\label{fig10} (Color online) Some states localized on the whole SW defects of a graphene.}
\end{figure}

The obtained results for three kinds of localized states are shown in Fig. 8, 9 and 10. The localized states
caused by pentagon defects are presented in Fig. 8, from which one can find their frequencies are very large,
and those modes with larger frequencies tend to be more localized on the pentagon rings. For the localized
states caused by heptagon defects shown in Fig. 9, the modes with larger frequencies also tend to be more
localized on the heptagon rings, but their largest frequency is about 1356 cm$^{ - 1}$, which is still lower
than the $G$ band frequency of graphene. All these phenomena agree qualitatively with our REBO results and the
discussions based on the simple model.

The localized modes in Fig. 10 can not be simply considered to be localized on which defects. In fact, they
distribute on the whole SW defect, which means they are caused by the whole SW defect. On the other hand, their
appearance also indicates that the distance between pentagon and heptagon rings has a great inference on the
behavior of localized states.

Quantitatively, the highest frequency of the localized states on the
pentagon rings in the SWNT IMJs should be higher than 1820 cm$^{ -
1}$ due to the curvature effect of nanotubes. In fact, in Ref.
\onlinecite{r37}, the authors studied the vibrational property of
the small diameter (3, 3) SWNT with SW defect and found a local
defect vibration mode with its frequency of 1962 cm$^{ - 1}$ due to
large curvature effect of the (3, 3) SWNT, which could be served as
a fingerprint of the SW defect in carbon nanotubes. From our
foregoing discussion, it can be deduced that this local defect
vibration mode is just the highest frequency localized mode mainly
caused by the pentagon defects.

Now, we want to study the relationship of the localized mode frequency with the defect positions, and so make
calculations on two structures, i.e., two (5, 5)-(10, 10) SWNT IMJs connected by different methods, whose
structures are shown in Fig. 11.

\begin{figure}
\includegraphics[width=3.24in,height=1.80in]{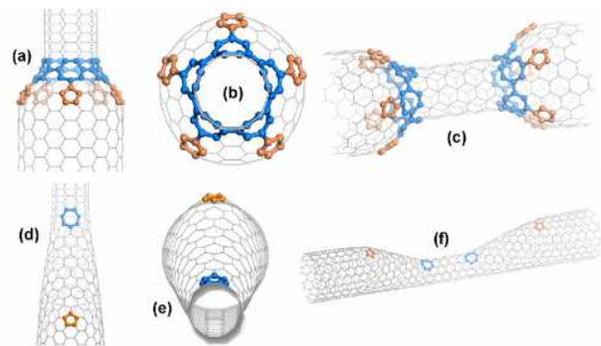}
\caption{\label{fig11} (Color online) Two different structures of the (5, 5)-(10, 10) SWNT IMJs, viewed from
different directions. (a), (b) and (c) are the high symmetry structures. (d), (e) and (f) are the low symmetry
structures.}
\end{figure}

The symmetry group of the first SWNT IMJ (Fig. 11a-11c) is $C_{5v} $, which is called as the HS (high symmetry)
SWNT IMJ. The symmetry group of the second SWNT IMJ (Fig. 11d-11f) is $C_s $, which is called as the LS (low
symmetry) SWNT IMJ. Beside their different symmetries, there are also another two important differences between
them, which are the junction length defined as the distance between the pentagon-heptagon defects, and the
number of pentagon-heptagon defects. There are five pairs of pentagon-heptagon defects in the HS SWNT IMJ, and
only one pair of pentagon-heptagon defects in the LS SWNT IMJ.

Now we will pay our attention only to the high frequency part of the Raman spectra of these two SWNT IMJs, which
have been shown in Fig. 12. It can be found from Fig. 12 that the $G$ band of the LS SWNT IMJ extends into a
wider region than that of the HS SWNT IMJ. This should root in the long transition region in the LS SWNT IMJ, in
which the IMJ radius and the bond lengths change almost continuously. So, most probably, the SWNT IMJ with a
longer junction region will have a wider $G$ band, which could be observed experimentally.

\begin{figure}
\includegraphics[width=3.24in,height=1.15in]{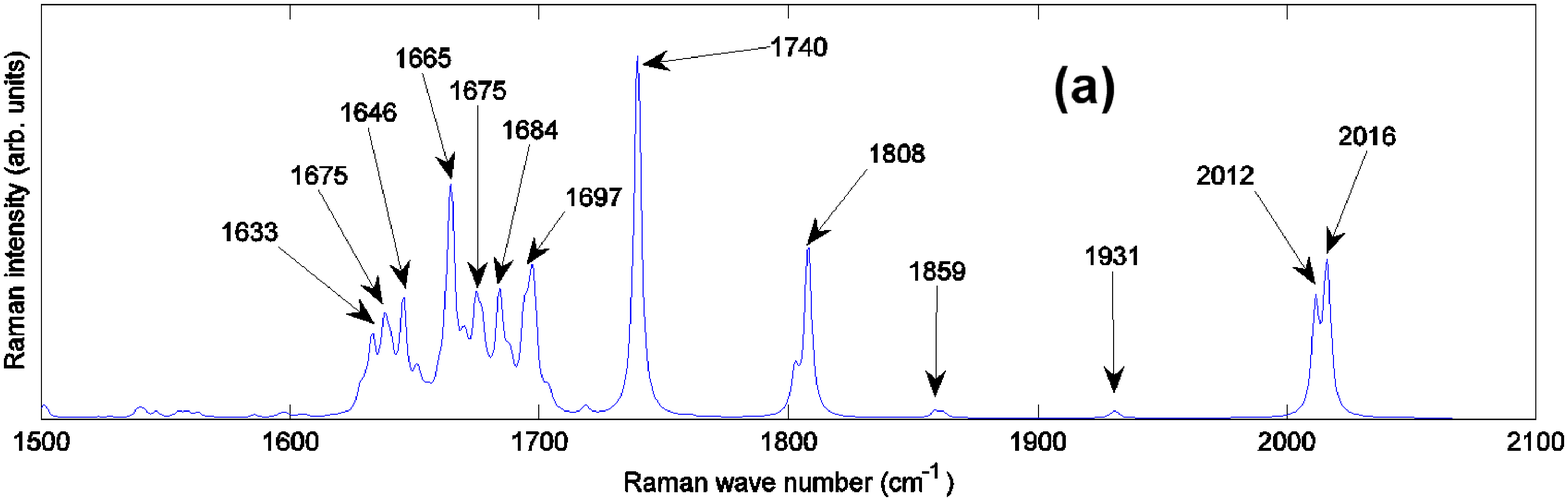}
\includegraphics[width=3.24in,height=1.15in]{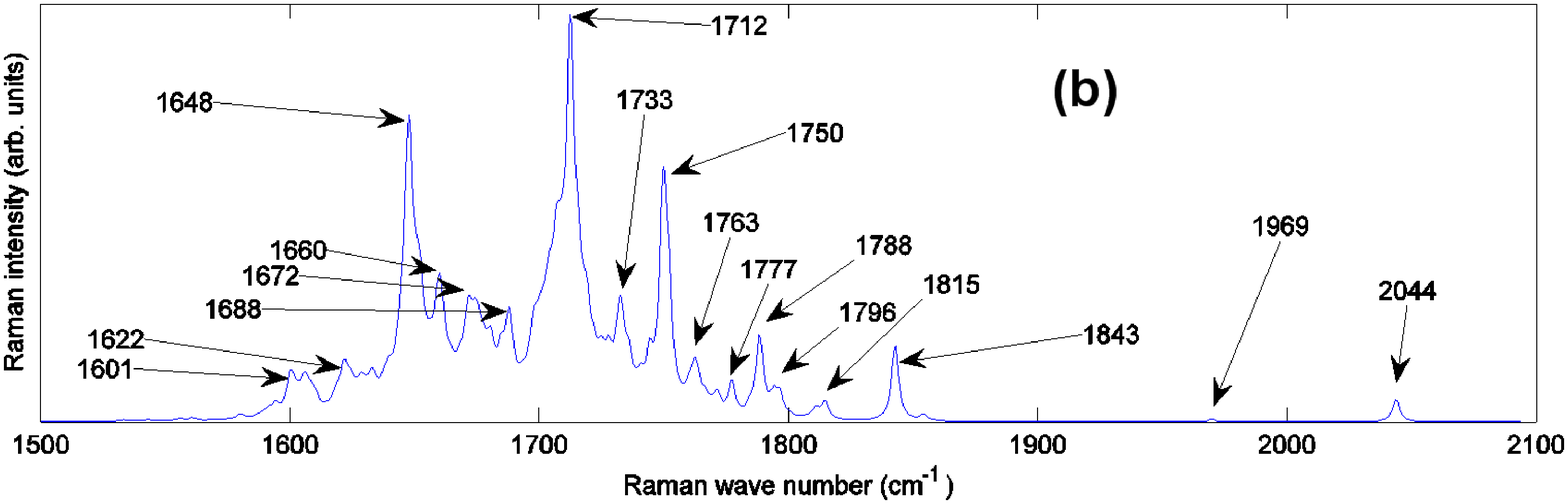}
\caption{\label{fig12} (Color online) The high frequency part of the Raman spectra of the two SWNT IMJs. The
upper and lower panels are for the high and low symmetry SWNT IMJ, respectively.}
\end{figure}

In addition, one can find that the intensities of the highest frequency localized modes in the two SWNT IMJs are
very different. In other words, relative to the $G$ band intensity, the highest frequency localized mode in the
HS SWNT IMJ has a larger intensity than that in the LS SWNT IMJ, which is probably due to the fact that there
exist five times more pentagon-heptagon defects in the HS SWNT IMJ than in the LS SWNT IMJ. So the intensity of
the highest frequency localized mode could be used to determine the number of pentagon rings experimentally.
Once the number of pentagon rings is determined, the number of heptagon rings can be determined too by Eq. 1.
Generally speaking, nice reference samples with a `known' number of pentagons will be greatly helpful to
determine the accurate number of defects in the IMJ, which is probably a technically difficult at the present
time. However, with improving gradually the experimental tools, the nice sample could be manufactured in future.
In a word, we think it is possible to use this method to determine the number of pentagon defects after we can
set up a standard for it by measuring the nice reference sample with a "known" number of pentagons, e.g., a
perfect SWNT with a cap at its end, which has six pentagons, or the SWNT with a Stone-Wales defect, which is
composed of two pentagon-heptagon pairs next to each other. At least, probably this method could give an
approximate number of pentagons in the SWNT IMJ. Finally, we also hope that it can be combined with other
experimental methods, e.g., the optical responses to reach more practically this goal.

\section{Conclusions}

In this work, the vibrational modes of some SWNT intramolecular
junctions have been calculated by using the newest Tersoff-Brenner
potential. Furthermore, using the empirical bond polarizability
model, the nonresonant Raman spectra of these SWNT IMJs have been
calculated. It is found that the highest frequency localized states
caused by pentagon defects could be used to identify accurately
detailed geometrical structure of the SWNT IMJ. That is because its
frequency lies out of the $G$ band of SWNTs and can be distinguished
easily. And its intensity has a close relation with the number of
pentagon defects. On the other hand, it is found that the heptagon
defects can only cause weak localized states below the $G$ band of
SWNTs and can hardly be observed in experiments. All these results
could be qualitatively explained by a simple ``topological
impurity'' model. The first-principles calculations are also carried
out to further prove our results quantitatively, and show that the
characteristic frequency of the SW defects should lie between 1820
and 1962 cm$^{ - 1}$, depending on the local curvature of the tube.
Furthermore, the junction length is shown to have an important
effect on the $G$ band width of the SWNT IMJ, i.e., the shorter
transition region will make the $G$ band of SWNT IMJ to be narrower
and simpler. Finally, the intensity of the Raman spectra has a close
relation with the number of topological defects.

\begin{acknowledgments}
The authors acknowledge support from the Natural Science Foundation
of China under Grant No. 10474035 and No. 90503012, and also support
from a Grant for State Key Program of China through Grant No.
2004CB619004.
\end{acknowledgments}

\end{document}